\documentclass[10pt,letterpaper]{article}
\usepackage{opex3}

\usepackage{graphicx}
\usepackage{dcolumn}
\usepackage{bm}
\usepackage{subfigure}
\usepackage{cite}

\newcommand{\ket}[1]{\ensuremath{\left|#1\right\rangle}}

\begin{document}

%% START HERE
%%%%%%%%%%%%%%%%%% title page information %%%%%%%%%%%%%%%%%%
\title{Demonstration of images with negative group velocities}

\author{Ryan T. Glasser$^{\ast}$, Ulrich Vogl, and Paul D. Lett}

\address{Quantum Measurement Division, National Institute of Standards and Technology, and\\
Joint Quantum Institute, NIST and the University of Maryland, Gaithersburg, MD 20899 USA\\}

\email{$^{\ast}$rglasser@nist.gov} %% email address is required



\begin{abstract}We report the experimental demonstration of the superluminal propagation of multi-spatial-mode images via four-wave mixing in hot atomic vapor, in which all spatial sub-regions propagate with negative group velocities.  We investigate the spatial mode properties and temporal reshaping of the fast light images, and show large relative pulse peak advancements of up to 64\,$\%$ of the input pulse width.  The degree of temporal reshaping is quantified and increases as the relative pulse peak advancement increases.  When optimized for image quality or pulse advancement, negative group velocities of up to $v_{g}=-\frac{c}{880}$ and $v_{g}=-\frac{c}{2180}$, respectively, are demonstrated when integrating temporally over the entire image.  The present results are applicable to temporal cloaking devices that require strong manipulation of the dispersion relation, where one can envision temporally cloaking various spatial regions of an image for different durations.  Additionally, the modes involved in a four-wave mixing process similar to the present experiment have been shown to exhibit quantum correlations and entanglement.  The results presented here provide insight into how to tailor experimental tests of the behavior of these quantum correlations and entanglement in the superluminal regime.\end{abstract}

 % REPLACE WITH CORRECT OCIS CODES FOR YOUR ARTICLE

%%%%%%%%%%%%%%%%%%%%%%% References %%%%%%%%%%%%%%%%%%%%%%%%%

%%%%%%%%%%%%%%%%%%%%%%%%%%  body  %%%%%%%%%%%%%%%%%%%%%%%%%%
\section{Introduction}
Optical pulse propagation with group velocities larger than the speed of light in vacuum, $c$, or negative, have been demonstrated theoretically and experimentally in a variety of systems \cite{garrett1970,chu1982,segard1985,segard2005,milonni2002,2010review}.  The anomalous dispersion required for generating ``fast" light occurs near the center of absorption lines and on the wings of gain lines \cite{bolda1994,boyd2003,novikova2004,gisin2004,thevenaz2005,chiao1994,dogariu2000}.  It has recently been shown that the seed and conjugate pulses involved in the four-wave mixing (4WM) process in hot rubidium vapor may exhibit large negative group velocities \cite{glasser2012}.

The group velocity $v_{g}$ is connected to the slope of the frequency dependent index of refraction $n(\nu)$ by:
\begin{eqnarray}\label{1}
v_{g}=\frac{c}{n_{g}}=\frac{c}{n(\nu)+n(\nu)\frac{dn(\nu)}{d\nu}},
\end{eqnarray}
where $n_{g}$ is the group index and $\nu$ is the optical frequency.  Operationally, the group velocity may be identified with the propagation speed of the peak of an optical pulse \cite{gauthier2007}.  Pulses propagating through a dispersive medium experience a relative delay of $\Delta T=\frac{L}{v_{g}}-\frac{L}{c}$, where $L$ is the length of the medium \cite{dogariu2000}.  When the group velocity is negative, $\Delta T$ is also negative, corresponding to a relative advancement of the pulse after traveling through the medium. It is well established (cf. \cite{segard2005,milonni2002,2010review}), that while dispersive media can alter the group velocity of a pulse, no superluminal transfer of information can be achieved.
In the present experiment we demonstrate the advancement of two-dimensional images carried by optical pulses that propagate through a region of anomalous dispersion, which is the complementary fast light analogue of the slow light image experiments performed by Camacho, et al. \cite{camacho2007}.  Using the inherently multi-spatial-mode 4WM process, we investigate the spatial variations of the group velocity and relative advancement of an image propagating with a negative group velocity.  A relative pulse peak advancement of 64\,$\%$ of the input pulse full-width at half-maximum (FWHM) is shown.  Additionally, we analyze the degree of temporal pulse reshaping as the relative pulse peak advancement is varied.
%Some spatial sub-regions of the pulses used in our experiment exhibit pulse advancements of $>100$\,ns,

The ability to impart images on an optical pulse propagating through a region of anomalous dispersion has a number of interesting applications.  Similar 4WM double-lambda schemes in rubidium have been shown to exhibit multi-spatial-mode entanglement \cite{lett2007}.  By sending one of the two entangled images through the present fast light medium, the spatial properties of the multi-spatial-mode entanglement propagating through anomalously dispersive media may be investigated.  Additionally, by spatially controlling the anomalous dispersion and relative pulse advancement, one can envision combining the present fast light system with a similar slow light system to develop a multi-spatial-mode temporal cloak analogous to the single-spatial-mode system in \cite{fridman2011}.

\begin{figure}[t]
\centering\includegraphics[width=8.4cm]{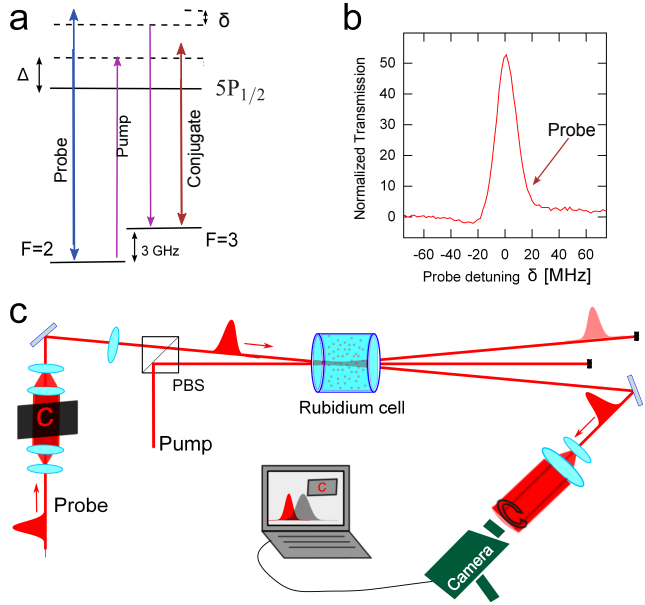}% Here is how to import EPS art
\caption{\label{fig:epsart} Schematic showing (1a) the double-lambda level scheme, (1b) the typical probe gain lineshape with probe detuning indicated, and (1c) the experimental setup.  The pump beam is detuned $\approx$\,400\,MHz to the blue of the $^{85}$Rb D1 line, with the probe blue-detuned $\approx$\,3.0\,GHz relative to the pump.  The probe's center frequency is set to be on the blue wing of the probe gain line.  After the 4WM interaction, the probe pulses are imaged onto a gated, intensified CCD camera.  The pulses are able to be time-resolved down to 2.44\,ns temporal bins.}
\end{figure}

\section{Experimental setup}
The 4WM process used here is pumped with a linearly-polarized continuous-wave laser detuned $\Delta\,\approx$\,400\,MHz to the blue of the Rb D1 line at $\lambda\,\approx\,795$\,nm, $\ket{5S_{1/2},F=2}\rightarrow\ket{5P_{1/2}}$, as shown in Fig.\,1(a).  The pump beam is produced by spatially filtering the output of a semiconductor tapered amplifier that is seeded with a tunable, single-mode diode laser.  The probe beam is generated by double-passing a 1.5\,GHz acousto-optic modulator (AOM) with light split off from the output of the same diode laser that produces the pump.  The probe beam is then pulsed by sending 200\,ns electronic square pulses into a second AOM with a $1/e$ rise time of $\approx\,190$\,ns, resulting in nearly Gaussian FWHM 200\,ns optical pulses.  The probe and pump beams are orthogonally polarized and combined on a polarizing beam splitter, at an angle of $\approx$\,1$^{\circ}$.  A conjugate pulse is created via the interaction and propagates at an angle satisfying the phase-matching condition. While previous experiments have shown that this conjugate pulse may also propagate superluminally \cite{glasser2012}, we focus here only on the probe mode and characterize the influence of the anomalous dispersion medium on the transverse modes of the pulse.  The probe is tuned such that its center frequency is on the blue wing of the gain line peak as indicated in Fig.\,1(b), in order to produce maximal relative pulse peak advancement.

\begin{figure}[t]
\centering\includegraphics[width=7.5cm]{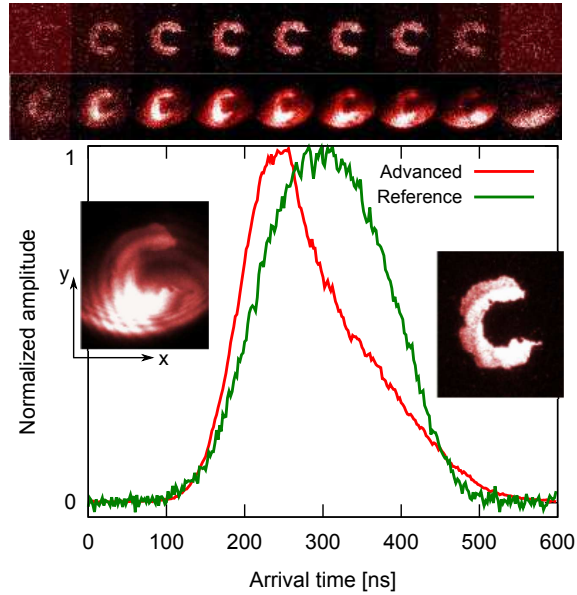}% Here is how to import EPS art
\caption{\label{fig:epsart} Negative group velocity of a carrier pulse resulting in an advanced pulse peak with a spatially
multi-mode image, in this case the letter ``c". The green curve
is the detected probe pulse, integrated over the image, when the pump is not present.
The red curve is the detected superluminal amplified probe pulse, integrated over the image, when
the pump is turned on. A relative advancement of $\approx$\,50 ns, corresponding to a group velocity of $v_{g}=-\frac{c}{880}$, is
shown. The probe pulse in this measurement was shaped with
the letter "c", and the arrival time was monitored with a gating
width of 3\,ns. The snapshots across the top of the graph
show the cross section of the beam at equidistant times between
120\,ns and 480\,ns (top row: reference, lower row: superluminal
pulse). The two insets show the full time-integrated
images. The advanced image (left inset) shows distortion due
to inhomogeneous gain, Kerr-lensing and leaked pump light,
but the principal shape clearly persists. The superluminal
pulse group velocity can be determined pixel-wise for the image,
as well as integrated over the whole image. The peak gain of the unnormalized superluminal pulse is $\approx$\,2.1.}
\end{figure}

The $220$\,mW pump beam is focused into the cell, resulting in an elliptical focal spot size of $\approx\,750\,\mu$m x $950\,\mu$m at the center of the cell.  The input probe beam is lightly focused into the cell with a nearly Gaussian FWHM spot size of $\approx\,600\,\mu$m, and a peak power of 10\,$\mu$W maximum.  The $^{85}$Rb cell is 1.7\,cm long and kept at a constant temperature of $115\,^{\circ}$C, corresponding to a number density of $\approx1.2\times10^{13}\,$cm$^{-3}$.  After the 4WM interaction, the probe pulses are detected with a gated, intensified CCD camera.  The camera allows us to examine $\approx$\,2.5\,ns time slices through the temporal envelope of the incident pulses, on each pixel.   Reference pulses are taken with the pump beam blocked, and correspond to pulses propagating at the speed of light in vacuum $c$, to within our experimental uncertainty (that is, the arrival time of reference pulses propagating along the optical path through the cell with no pump beam present and those propagating along the optical path when the cell is removed show no arrival time difference to within our experimental uncertainty).

\begin{figure}[t]
\centering\includegraphics[width=7.4cm]{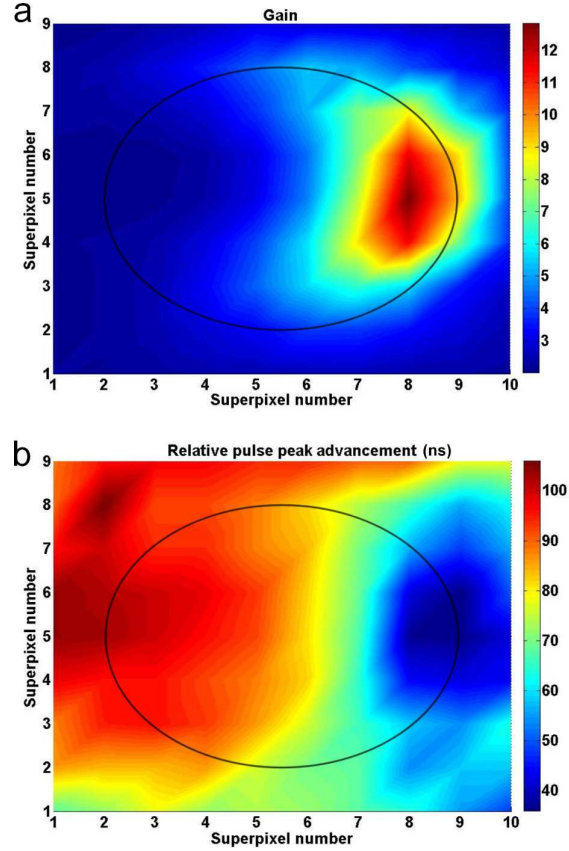}
\caption{\label{fig:wide} Gain (3a) and relative pulse peak advancement (3b) of the superluminal probe pulses for the input probe with a Gaussian spot.  Each superpixel corresponds to a 12$\times$12 binning of pixels on an intensified CCD camera.  The pump beam is slightly elliptical, with a waist of $\approx750\,\mu$m$\times950\,\mu$m.  Relative pulse peak advancement is seen to increase from $\approx$\,40\,ns to $\approx$\,100\,ns from the right-hand side to the left-hand side of the probe spot.  The gain also varies spatially, with the highest gain regions corresponding to the lowest relative pulse peak advancements.  The ellipses correspond to the $1/e^{2}$ intensity of the detected amplified probe spots.  Uncertainties in the relative pulse peak advancement are largest toward the edges of the image ($\approx$\,10\,ns), due to statistical uncertainties from a decreased signal-to-noise ratio resulting from the lower intensities.  The uncertainty in the relative advancement near the inner region of the image is $\approx$\,3\,ns, resulting from the minimum detector gating time.}
%\end{center}
\end{figure}

\section{Results}
Due to the inherently multi-spatial-mode nature of the 4WM without the presence of a cavity, the probe pulse is able to carry an image throughout the process.  In Fig.\,2, we show an image of a ``c" on the probe pulse.  The amplified probe pulse exhibits a relative pulse peak advancement of $\approx$\,50\,ns on a 200\,ns optical pulse, when integrated over the entire image, corresponding to a group velocity of $v_{g}=-\frac{c}{880}$.  Smaller subsections of the image all propagate with negative group velocities, with advancements varying from $\approx$\,10\,ns to $>$\,50\,ns.  The superluminal image is distorted, but is clearly visible.  The current experimental limitations that result in degraded image quality are primarily the available pump and probe powers.  With sufficient powers, it should be possible to enlarge the pump and probe beams such that they are collimated and nearly uniform across the interaction region, resulting in equal advancements across the image. The relative pulse peak advancements are found by taking the peak of the output pulse and comparing its arrival time to the arrival time of the peak of a pulse propagating at $c$.  The uncertainty of the pulse peak advancement in the outer regions of the images is due primarily to the difficulty of determining the pulse peak due to low optical power in these regions, and is roughly 10\,ns.  The uncertainty of the pulse peak advancement near the inner regions of the images where higher intensities are present is limited by the minimum detector gating time, and is $\approx\,3$\,ns.  %       All uncertainties shown are one standard deviation, with the major contributions to the uncertainties in the outer and inner regions of the image being statistical and the minimum detector gating time, respectively.

In order to better analyze the spatial properties of superluminal images, we use a nearly Gaussian spatial image. The probe beam's center frequency is set at the point of optimal advancement of the integrated spatial profile of the pulses.  If the probe and pump beams were perfectly collimated with a uniform intensity, all spatial regions of the probe are expected to be advanced uniformly.  In order to create a spatially varying group velocity profile, one can take advantage of angular dispersion \cite{torres2010}, which allows for the translation of a phase-mismatch of the probe and pump beams into a spatially varying group-index \cite{Kumar1994}.  To accomplish this we adjust the probe focus so that the beam waist crosses the pump beam slightly off-center.  Due to the strong dispersion of the gain line, as outlined in \cite{Kumar1994}, the phase-matching condition is fulfilled for a spread of angles determined by $\sqrt{\lambda/L}$ (where $\lambda$ is the wavelength and $L$ is the length of the medium), corresponding to 7\,mrad ($\approx$\,0.4$^{\circ}$) in our case.  The k-vectors of the pump and probe beams now intersect with a varying angle across the  interaction volume in the cell, so that the phase-matching changes across the beam.
 As seen in Fig.\,3(a), this results in a spatial gradient of the gain. Additionally, this results in the spatially-varying group index profile shown in Fig.\,3(b).
In the data shown the k-vectors follow a nearly linear gradient in the horizontal direction and the observed change in group advancement $\Delta T(x)$ follows the relation $\Delta T(x) \sim x \frac{\partial}{\partial x}(\frac{\partial k(x,\omega)}{\partial \omega})$, where $x$ denotes the horizontal transverse beam direction \cite{torres2010}.
 Data is shown as a function of superpixel position, which are pixels binned into 12$\times$12 groups.  This provides more light on each superpixel and increases the signal-to-noise (the ellipses in the figures show the approximate $1/e^{2}$ intensity of the amplified probe spots).  Spatially-varying gain averaged over individual superpixels ranges from $\approx$\,2 to $\approx$\,12. Relative pulse peak advancements range from $\approx$\,40\,ns to $\approx$\,95\,ns.  This demonstrates another degree of freedom that can be used to control the relative pulse peak advancement.  In principle, one can engineer how the probe is focused into the cell to produce different spatially-varying pulse advancements across the image.

%\begin{figure*}[t]
%\begin{center}
%\centering\subfigure{
%\includegraphics[scale=0.293]{gain2d.eps}}
%\centering \subfigure{
%\includegraphics[scale=0.293]{advancement2d.eps}}

Having analyzed some of the spatial properties of the superluminal pulses, we now turn to the temporal profile.  The amount of temporal reshaping that a superluminal pulse experiences in general gets larger with increasing advancement.  To characterize this distortion, we employ a metric similar to that used in \cite{bigelow2006}, as it is a convenient measure to use in an experimental setting without the need for employing phase-sensitive measurements.  The degree of distortion is defined as:
\begin{equation}\label{2}
D=\sqrt{\int^{\infty}_{-\infty}\left|\frac{|E'(z+L,t)|^{2}}{\int|E'(z+L,t)|^{2}dt}-\frac{|E(z,t-\Delta T)|^{2}}{\int|E(z,t-\Delta T)|^{2}dt}\right|dt}.
\end{equation}
Here $E'$ and $E$ are the output and reference pulse envelopes, respectively.

To quantify the temporal pulse reshaping, we use Eq.\,(2) to analyze the measured pulses after normalization.  We choose Eq.\,(2) to quantify reshaping that is due only to propagation through the region of anomalous dispersion as it is zero for identically shaped pulses, even if gain is present, and is nonzero for any reshaping.  This is similar to other metrics for measuring distortion, although others may purposely not normalize the advanced and reference pulses, and thus are nonzero when only gain is present \cite{segard2005}.  In order to see the effect of relative advancement on the pulse distortion, we vary the relative pulse peak advancement from $\approx$\,5\,ns to $\approx$\,75\,ns by changing the input probe power as in \cite{glasser2012}.  The degree of pulse reshaping increases over this range of advancements from D\,$\approx$\,0.45 to D\,$\approx$\,0.6.  The principle reshaping can best be described as a narrowing of the pulse, with a steeper rising edge than falling edge, albeit with some ringing on the falling edge at the largest advancements.  Reshaping results from contributions of the higher order terms in the expansion of $n(\nu)$, such as group velocity dispersion \cite{gauthier2007}.  The rising edges are advanced by a smaller amount than the falling edges, consistent with the analysis performed in \cite{macke2003}.  Additionally, the pulse temporal advancement varies spatially across the probe spot as shown in Fig.\,3(b).  The fundamental temporal reshaping of the pulse, however, is relatively uniform across the spot.  For the 2$\times$2 fastest binned spot, the rising and falling edges at FWHM of the 200\,ns pulses are advanced by $\approx\,38$\,ns and $\approx\,156$\,ns, respectively, and displays a pulse peak advancement of $\approx$\,100\,ns.  Integrating over the entire probe spot, the rising and falling edges at FWHM are advanced by $\approx\,24$\,ns and $\approx\,124$\,ns, respectively, and an integrated pulse peak advancement of $\approx$\,80\,ns is measured. This relative advancement is rather large considering the amount of peak gain and absorption theoretically required to achieve comparable advancements in a similar but somewhat constrained system, where a constant background gain of $~$e$^{64}$ and an absorption at the line-center of $~$e$^{-32}$ is required to achieve a relative advancement of $2\sqrt{2}$ times the input pulse width\cite{Narum2010}.

\begin{figure}[t]
\centering\includegraphics[height=6.0cm]{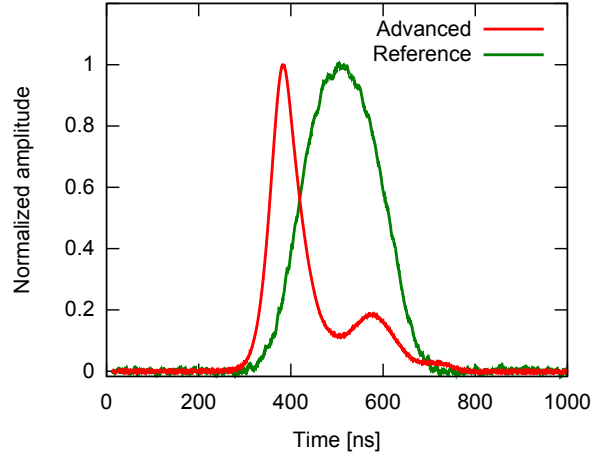}% Here is how to import EPS art
\caption{\label{fig:epsart} Plot of the advanced pulse versus time when the system is optimized for maximum advancement rather than image quality.  The red and green curves are the advanced pulse and reference pulse intensities integrated over the entire Gaussian spot.  A pulse peak advancement of 124\,ns is shown, corresponding to a relative pulse peak advancement of 64\,$\%$ compared to the input pulse FWHM, and a group velocity of $v_{g}=-\frac{c}{2180}$.  The gain in this case is $\approx$\,5, and the relative degree of reshaping is D\,$\approx$\,0.8.  Ringing on the trailing edge of the advanced pulse is seen, as expected for very large relative advancements.}
\end{figure}

According to \cite{macke2003}, the rising and falling edges of a temporally Gaussian pulse will be advanced, respectively, by:
\begin{eqnarray}\label{2}
T_{adv,\uparrow}=T_{adv}-\tau_{a}+\frac{\tau_{a}}{\beta}\\
T_{adv,\downarrow}=T_{adv}-\tau_{a}-\frac{\tau_{a}}{\beta}.
\end{eqnarray}
Here $T_{adv,\uparrow}$, $T_{adv,\downarrow}$ and $T_{adv}$ are the advancement of the leading edge, the advancement of the falling edge and the center-of-gravity advancement of the pulse (which we take to be the pulse peak for consistency with the analysis above), $\tau_{a}$ is the half-width of the pulse at the point where the values are calculated and $\beta$ is the factor by which the intensity profile of the pulse is narrowed.  Analyzing these values at the FWHM of the pulses, we find for the total Gaussian spot for the rising and falling edges, $\beta_{\uparrow}=1.67$ and $\beta_{\downarrow}=2.04$, respectively.  We also examine the region of the Gaussian spot that exhibits the largest advancement, the 2$\times$2 area of superpixels (1,5) to (2,6) in Fig.\,3.  Similar analysis for the 2$\times$2 binned spot mentioned above results in $\beta_{\uparrow}=2.3$ and $\beta_{\downarrow}=2.68$.  As expected, the 2$\times$2 binned spot exhibits more significant reshaping accompanying the larger relative pulse advancement.  In all cases, the falling edge is advanced more than the rising edge, as expected \cite{macke2003}.

We are able to increase the advancement, at the cost of increased pulse and image distortion, by decreasing the size of the probe and pump spots in the cell.  Figure 4 shows such a case, where a relative pulse peak advancement of $>$60$\,\%$, compared to the input pulse FWHM, is obtained using a gain of $\approx 5$.  This corresponds to a group velocity of $v_{g}=-\frac{c}{2180}$.  The degree of pulse reshaping is D\,$\approx$\,0.84, and ringing after the main peak is seen, as theoretically predicted \cite{macke2003}.  The pulse is again narrowed, with rising and falling edge advancements of $\approx\,60$\,ns and $\approx\,190$\,ns, respectively, and a pulse peak advancement of $\approx$\,124\,ns on a 200\,ns pulse.  This is, to our knowledge, the largest relative pulse peak advancement of an optical pulse demonstrated experimentally.  While the advancement demonstrated here is significant compared to the best previous relative pulse advancement (to our knowledge) of 42\,$\%$ \cite{segard1985,2010review}, in a molecular absorption system, it exhibits larger distortion and less advancement when compared to a system with an optimized transfer function \cite{segard2005}.  As shown in \cite{segard2005}, a fast light system with an optimal transfer function can allow for 100\,$\%$ relative pulse advancement with a peak gain of 84, and a distortion of 15\,$\%$ using a similar definition to Eq.\,(2).  Limitations to the relative pulse peak advancement in the present scheme are primarily due to the pump and probe powers.  Additionally, limitations resulting in image distortion through the region of anomalous dispersion are due primarily to phase-matching constraints, over which there is a finite allowable bandwidth where appreciable 4WM gain can take place.  The group velocity dispersion in the present system is determined by the lineshape resulting from the 4WM process.  The bandwidth of anomalous dispersion resulting from the 4WM gain line limits the usable probe pulse widths to being larger than roughly 75\,ns.  Finally, noise added due to the phase-insensitive nature of the process should also be considered in future experiments if one is interested in investigating the behavior of quantum correlations in such a system.

\section{Conclusion}
We have experimentally demonstrated images propagating with negative group velocities.  We have investigated the spatial variation of the relative pulse peak advancement and gain on pulses that have negative group velocities due to anomalous dispersion resulting from the 4WM process.  The entirety of a nearly Gaussian spatial spot exhibits large negative group velocities in all spatial subregions. Three knobs may be tuned to vary the amount of relative pulse peak advancement, the pump power, the input probe power, and the k-vector of the probe relative to the pump.  Additionally, we have analyzed the degree of temporal reshaping that the pulses exhibit after having traversed the fast light medium.  Finally, we show relative pulse peak advancements of $>$60$\,\%$ relative to the 200\,ns FWHM input pulses, corresponding to a group velocity of $v_{g}=-\frac{c}{2180}$.  These results should prove to be beneficial when trying to engineer fast light systems to exhibit specific desired properties, such as the amount of gain, advancement and reshaping.

The flexibility of the present system should allow for investigations into the effects of negative group velocities on quantum correlations and squeezing, as well as implementations of a temporal cloak over multiple spatial modes.  The ability to vary the group velocity of optical pulses spatially is a step toward the demonstration of a spatially-varying temporal cloak.  One implementation of temporal cloaking utilizes a ``split time-lens," in which a pulse is effectively split in time to create a temporal gap where the original pulse resided \cite{fridman2011}.  The split pulse is then closed, and whatever event occurred in the time gap is hidden.  The ability to manipulate the group velocity to advance pulses by different amounts in different spatial regions as demonstrated here could allow the temporal cloaking of different regions of spatially multimode pulses by different durations, which is not possible when using single-mode fibers as in \cite{fridman2011}.  Additionally, our results are applicable to the investigation of the effects of superluminal propagation on bipartite entangled states.  A nearly identical 4WM setup to the one used here has shown that the probe and conjugate modes can exhibit quantum correlations and entanglement \cite{lett2007b,alberto2008}.  In analogy to the experiment in \cite{boyer2007}, it should be possible with the present setup to explore the behavior of quantum correlations under conditions when superluminal propagation occurs. By taking advantage of the spatially varying group index, one can measure the cross-correlation as a function of advancement, pixel-by-pixel.

\section*{Acknowledgments} This work was supported by the Air Force Office of Scientific Research.  This research was performed while Ryan Glasser held a National Research Council Research Associateship Award at NIST.  Ulrich Vogl would like to thank the Alexander von Humboldt Foundation.

\end{document}